\documentstyle[12pt]{article}
\newcommand{\beq}{\begin{equation}}
\newcommand{\eeq}{\end{equation}}
\newcommand{\la}[1]{\label{#1}}

\newcommand{\ep}{\varepsilon}
\begin{document}
\title{Calculations of Scattering Lengths in Four-Nucleon System on the 
Basis of Cluster Reduction Method for Yakubovsky Equations} 
\author{S.L.Yakovlev\thanks{E-mail: {\tt
yakovlev@snoopy.niif.spb.su}}, I.N.Filikhin}
\date{ }
\maketitle
\centerline{\small Department of Computational Physics, St.Petersburg State 
University} 
\centerline{\small 198904 St.Petersburg, Petrodvoretz, Uljanovskaya str.
1, RUSSIA}  

\abstract{
The cluster reduction method for the Yakubovsky equations in
configuration space is used for calculations of zero-energy scattering
in four-nucleon system. The main idea of the method
consists in making use of expansions for the
Yakubovsky amplitudes onto the basis of the Faddeev components
for the two-cluster
sub-Hamiltonian eigenfunctions. The expansions reduce the original
equations to ones for the functions depending on the relative
coordinates between the clusters. On the basis of the resulting
equations the
$N$--$(NNN)$  zero-energy scattering problems are 
solved numerically with the
MT~I--III model for $N$--$N$ forces and neglecting the Coulomb
interaction between protons. 
}
\section{Introduction}

In this paper we continue our investigations of the four particle scattering
problem making use the cluster reduction method for Yakubovsky equations
in configuration space formulated in \cite{yadfiz1,yakfil3}.
The Yakubovsky differential equations (YDE) 
\cite{merkyak}, being the direct generalization of the three body 
Faddeev approach, inherit all advantages of latter. Two main of
them are the very simple way to impose the boundary conditions
corresponding to all elastic and rearrangement channels and minimum 
inputs needed to formulate the equations and boundary conditions.

For the first time, YDE have been used for calculations of the four 
identical particles bound 
state in \cite{myg}. It was observed that even for the simple model $N-N$
interactions the converged results within the usual numerical
approximations  of the YDE could be obtained with supercomputer 
facilities only. So, 
\newpage 
\noindent the progress in four nucleon calculations with
YDE depends on a new numerical solution technique. For the bound state
problem such a technique was elaborated in \cite{kok} and allowed to
improve the results of \cite{myg} considerably. For the scattering 
problem one of the new solution method was introduced in
\cite{yadfiz1, yakfil3, yadfiz2}.

Here, we employ the method for calculations of zero-energy $N$--$(NNN)$
scattering neglecting the Coulomb interaction between
protons. This system could be considered as a model for $n$--$^3$He, 
$p$--$^3$H or $n$--$^3$H scattering depending on a value of the total
isospin.    
   
\section{Yakubovsky equations}

In the Yakubovsky approach, the four particle wave function $\Psi $ 
should be decomposed into components in one to one correspondence to 
all chains of partitions. The chains consist of two cluster $a_{2}$
({\it e.g.}, $(ijk)l$ or $(ij)(kl)$) and three cluster $a_{3}$ 
({\it e.g.}, $(ij)kl$) partitions obeying the relation $a_{3}\in a_{2}$. The
latter means that the partition $a_{3}$ can be obtained from
partition $a_{2}$ by splitting of one subsystem. It is easy to see
that there exist $18$ chains of partitions for the four particle
system. 

The Yakubovsky wave function components can be defined by formulas
\cite{merkyak,myg}
\beq
\Psi_{a_{3}a_{2}} = R_{a_{3}}(E^{+})V_{a_{3}}\sum_{(b_{3}\neq
a_{3})\in a_{2}}R_{0}(E^{+})V_{b_{3}}\Psi ,
\la{yakcomp}
\eeq 
where
$$
R_{a_{3}}(E) = (H_{0}+V_{a_{3}}-E)^{-1},\ R_{0}(E) = (H_{0}-E)^{-1},\
 E^{+} = E+i0,
$$  
and $V_{a_{3}}$ stands for the two particle potential acting inside
the two particle subsystem of a partition $a_{2}$. If the  function
$\Psi $ is the solution of the Schroedinger equation 
$$ 
(H_{0}+\sum_{a_{3}}V_{a_{3}}-E)\Psi = 0,
$$ 
then the components $\Psi_{a_{3}a_{2}}$ obey the Yakubovsky equations
\cite{merkyak,myg}
$$
(H_{0}+V_{a_3}-E)\Psi_{a_{3}a_{2}}+V_{a_3}\sum_{({c_3}\neq {a_3})\in 
{a_2}}\Psi_{c_{3}a_{2}} = 
$$  
\beq
-V_{a_3}\sum_{{d_2}\neq {a_2}}\sum_{({d_3}\neq {a_3})\in 
{a_2}}\Psi_{d_{3}d_{2}} . 
\la{yeq}
\eeq

There exists a remarkable rule which allows to construct the
Schroedinger equation solution from the components $\Psi_{a_{3}a_{2}}$
 \cite{merkyak,myg}
$$
\Psi = \sum_{a_2}\sum_{a_3}\Psi_{a_{3}a_{2}}.
$$ 

Note that, the only inputs needed to formulate the YDE (\ref{yeq}) 
are the interparticle potentials.     

The second advantage of the YDE consists in the structure of the boundary
conditions. In contrast to the total wave function for the multichannel 
scattering, the asymptotics of the Yakubovsky component has the one 
channel form. Namely, $\Psi_{a_{3}a_{2}}$ involves in the asymptotical 
region the characteristics of the bound states for subsystems of
the partitions  $a_{2}$ only. As a consequence, the asymptotic 
behavior of  $\Psi_{a_{3}a_{2}}$ can be described in terms of one set 
of the relative coordinates corresponding to the chain  $a_{3}a_{2}$.

Let us turn to description of this asymptotics. To this end introduce 
the relative Jacobi coordinates for the four particle system. 
There exist 18  
sets of Jacobi coordinates in one to one correspondence to 18 chains 
of partitions. Note, that among them only two sets are topologically 
different. One corresponds to chains with partitions $a_2$ of the 
type $3+1$ ({\it e.g.,} (123)4) and the other one does to chains 
with partitions
$a_2$ of the type $2+2$ ({\it e.g.,} (12)(34)). These coordinates can
be given by formulas  
\beq
\begin{array}{l}
{\bf x}_{a_3} = {\bf r}_{i} - {\bf r}_{j} \nonumber \\
{\bf y}_{{a_3}{a_2}} = ({\bf r}_{i}+{\bf r}_{j})/2 - {\bf r}_{k} \\
{\bf z}_{a_2} = ({\bf r}_{i}+{\bf r}_{j}+{\bf r}_{k})/3 - {\bf r}_{l} 
\nonumber
\end{array}
\la{jac1}
\eeq
for  $a_{2}=(ijk)l, a_{3}=(ij)kl$ and
\beq
\begin{array}{l}
{\bf x}_{b_3} = {\bf r}_{i} - {\bf r}_{j} \nonumber \\
{\bf y}_{{b_3}{b_2}} = {\bf r}_{k} - {\bf r}_{l} \\
{\bf z}_{b_2} = ({\bf r}_{i}+{\bf r}_{j})/2 - ({\bf r}_{k}+{\bf r}_{l})/2 
\nonumber
\end{array}
\la{jac2}
\eeq
for  $b_{2}=(ij)(kl), b_{3}=(ij)kl$.

In this paper we are considering only binary processes, it assumes
that only the two clusters channels are open. In this case, the
components which correspond to the initial state with binding
clusters of subsystems of the partition $l_2$
have the form
\beq
\begin{array}{l}
\Psi_{{a_3}{a_2}}({\bf X},{\bf p}_{l_2}) = \delta_{{a_2}{l_2}}\psi^{a_3}_{
a_2}({\bf x}_{a_2})\exp \{ i({\bf p}_{a_2},{\bf z}_{a_2})\} \nonumber \\
+\psi^{a_3}_{a_2}({\bf x}_{a_2})U_{{a_2}{l_2}}({\bf z}_{a_2},{\bf p}_{l_2})+
\delta U_{{a_2}{a_3,{l_2}}}({\bf X},{\bf p}_{l_2}).
\end{array}
\la{cas}
\eeq
Here $ {\bf x}_{a_2} = \{ {\bf x}_{a_3},{\bf y}_{{a_3},{a_2}}\}$, 
${\bf X} = \{{\bf x}_{a_2},{\bf z}_{a_2}\}$
and ${\bf p}_{l_2}$ is the relative momentum conjugated to the vector
${\bf z}_{l_2}$.
The momentum ${\bf p}_{l_2}$ and the energy of the system $E$ 
obey the equation 
\[
E = \ep_{l_2} + \delta_{l_2} \frac{\hbar^{2}}{m}({\bf p}_{l_2})^2 . 
\]
 $\ep_{a_2}$ and $\psi^{a_3}_{a_2}$ are binding energies and the Faddeev 
components of the bound  states for the two cluster Hamiltonians
\[
h_{a_2} = h_{a_2}^{0}+\sum_{{a_3}\in {a_2}}V_{a_3} =
-\frac{\hbar^{2}}{m} (\Delta_{{\bf x}_{a_3}}+
\delta_{a_2} \Delta_{{\bf y}_{{a_3}{a_2}}})+\sum_{{a_3}\in {a_2}}V_{a_3}(
{\bf x}_{a_3}),
\]
where $\delta_{a_2} = 3/4 $ for the partitions $a_2$ of the type (3+1) and 
$\delta_{a_2} = 1$
for the partitions $a_2$ of (2+2) type. 

Amplitudes $U$ become the spherical waves 
\beq
U_{{a_2}{l_2}}({\bf z}_{a_2},{\bf p}_{l_2}) \sim {\cal A}_{{a_2}{l_2}}
\frac{\exp \{ i\sqrt{E-\ep_{a_2}}|{\bf z}_{a_2}|\} }{|{\bf z}_{a_2}|}
\ \ (|{\bf z}_{a_2}|\rightarrow \infty ).
 \la{as6}
\eeq
Amplitudes $\delta U_{{a_2},{a_3},{l_2}}$ are exponentially decreasing
functions as $|{\bf X}|\rightarrow \infty$ corresponding to virtual
breakup processes.

\section{ Cluster reduction of YDE.}

The cluster reduction of YDE \cite{yakfil3} consists in expansion of the
components $\Psi_{{a_3}{a_2}}$ onto the basis consisting of functions 
depending on the coordinates  ${\bf x}_{a_2}$. The most suitable 
basis for that is the one of the eigenfunctions for the operators
from the left hand side of the YDE (\ref{yeq}) 
\beq
(h_{{a_2}}^{0} + V_{a_{3}})\psi_{{a_2},k}^{a_3} + V_{a_3}
\sum_{({c_3}\neq {a_3})\in {a_2}}\psi_{{a_2},k}^{c_3} = 
\ep_{a_2}^{k}\psi_{{a_2},k}^{a_3}. 
\la{feqbas}
\eeq

The expansion of the components  has the form 
\beq
\Psi_{{a_3}{a_2}} = \sum_{k=0}^{\infty}\psi_{{a_2},k}^{a_3}({\bf x}_{a_2})
F_{a_2}^{k}({\bf z}_{a_2}), 
\la{expcom}
\eeq 
where the coefficients $F_{a_2}^{k}({\bf z}_{a_2})$ depend on the relative 
coordinates between the
clusters only.

The basis of the solutions of the eqs. (\ref{feqbas}) is 
not an orthogonal one, because the eqs. (\ref{feqbas}) are not
Hermitean. Fortunately, it is easy to show that the Hermitean
conjugated equations to the eqs. (\ref{feqbas}) are of the form
\cite{yaktmf} 
\beq
(h_{{a_2}}^{0} + V_{a_{3}})\phi_{{a_2},k}^{a_3} + 
\sum_{({c_3}\neq {a_3})\in {a_2}}V_{c_3}\phi_{{a_2},k}^{c_3} = 
\ep_{a_2}^{k}\phi_{{a_2},k}^{a_3}.
\la{confeqbas}
\eeq 
In the papers \cite{yaktmf} it is shown that equations (\ref{feqbas})
and (\ref{confeqbas}) allow the solutions of two classes. They are 
the physical ones, {\it i.e.,} the solutions of eqs. (\ref{feqbas})
which sum 
$$
\sum_{{a_3}\in {a_2}} \psi_{{a_2},k}^{a_3} = \psi_{{a_2},k}
$$ 
is the solution of the Schroedinger equation 
$$
h_{a_2} \psi_{{a_2},k}= \ep_{a_2}^k \psi_{{a_2},k},
$$
and the solutions of eqs. (\ref{confeqbas}) with the property
$$
\phi_{{a_2},k}^{a_3} =  \psi_{{a_2},k}
$$
for any ${a_3}\in {a_2}$. The second class of solutions of eqs.
(\ref{feqbas})  is formed by the spurious solutions for which 
$$
\sum_{{a_3}\in {a_2}} \psi_{{a_2},k}^{a_3} = 0.
$$
Unfortunately, there is no a such simple formula for the spurious
solutions of eqs. (\ref{confeqbas}). It is important to note, that
the complete basis is formed only by the physical and spurious
solutions together. Complete bases of solutions of eqs.
(\ref{feqbas}) and (\ref{confeqbas}) are biorthogonal
$$ 
\sum_{{a_3}\in {a_2}} \langle
\phi_{{a_2},l}|\psi_{{a_2},k}^{a_3}\rangle  = 
\delta_{lk}.
$$
This property allows to apply the projection technique in order to
obtain the equations for coefficients $F_{a_2}^{k}$ from series 
(\ref{expcom}) after substitution of eqs. (\ref{expcom}) into 
eqs. (\ref{yeq}). The result reads
$$
(-\frac{\hbar^{2}}{m}\delta_{a_2} \Delta_{{\bf z}_{a_2}} - E + 
\ep_{a_2}^{k})F_{a_2}^{k}(
{\bf z}_{a_2}) =   
$$
\beq
-\sum_{{a_3}\in {a_2}}\langle 
\phi_{a_2,k}^{a_3}|V_{a_3}\sum_{{d_2}\neq {a_2}}\sum_{({d_3}\neq
{a_3})\in {a_2}}\sum_{l\geq
0}\psi_{{d_2},l}^{{d_3}}F_{d_2}^{l}({\bf z}_{d_2})\rangle ,
\la{redyeq}
\eeq  
where the brackets $\langle . |. \rangle $
stand for the integration over ${\bf x}_{a_2}$.

The asymptotic boundary conditions for the coefficients
$F_{a_2}^{k}({\bf z}_{a_2})$ can be obtained from (\ref{cas}), (\ref{as6})
 by projecting and have the
following 'two body' form
$$
F_{a_2}^{0}({\bf z}_{a_2}) \sim \delta_{{a_2}{l_2}}\exp
{i({\bf p}_{a_2},{\bf z}_{a_2})} 
+{\cal A}_{{a_2}{l_2}}|{\bf z}_{a_2}|^{-1}\exp{i\sqrt{E-\ep_{a_2}}|{
\bf z}_{a_2}|}
$$
for the open channels and
$$
F_{a_2}^{k}({\bf z}_{a_2}) \sim 0 \ \ , k\geq 1
$$
for the closed channels.

The equations (\ref{redyeq}) are the desired coupled channel
equations for the two cluster collisions in the four body system.

\section{Application to the $N-NNN$ scattering problem.}

The eqs. (\ref{redyeq}) can be applied to solving the $N-NNN$
scattering problem after a suitable partial wave analysis. 
We have used the MTI-III model for the $N-N$
forces and the isotopic spin formalism. Within this conditions and 
neglecting the Coulomb interaction the channels corresponding to total
value of isospin T are independent. 

The calculated values of scattering lengths in $N-NNN$ system 
for a given values of the total spin $S$ and the total isospin $T$ 
are given in the Tab. 1
with data of direct solution of YDE from the paper \cite{carbon}. 
As one can see the agreement of calculations is rather reasonable.
In the Tab. 2 we collect the available data for $n-{^3}$H ($T=1$) 
scattering lengths calculated within various approaches.
As one can see the agreement of the calculations is quite good.
By the end in the Tab. 3, we give the results of calculations of scattering 
lengths in the channel $T=0$ performed under conditions when 
the spurious part of basis functions is removed from expansions
(\ref{expcom})
in order to show how the presence of the spurious solutions affects 
on the results of calculations. 

\section*{Acknowledgments}
This work was supported by Russian Foundation for Basic Research 
Grant No.~96-02-17021.

\newpage 
\vskip 5cm 
\hspace*{50mm} Tab. 1 \\ 
\begin{center}
\begin{tabular}{|c|c|c|c|}
\hline 
$S$   &  $T$   &  \cite{carbon}& present work \\
\hline 
0   &   0   &    14.75      & 14.7    \\
\hline
1   &   0   &    3.25       & 2.9      \\
\hline
0  & 1      &     4.13      & 4.0      \\ 
\hline 
1  &   1    &     3.73      & 3.6      \\
\hline 
\end{tabular}
\end{center}
\newpage 
\vskip 5cm 
\hspace*{50mm} Tab. 2\\ 
\begin{center}
\begin{tabular}{|c|c|c|}
\hline
\parbox{4cm}{\center Refs.} & \parbox{2cm}{\center $A_{S=0}$ fm} & 
\parbox{2cm}{\center $A_{S=1}$ fm} \\
             &           &          \\ 
\hline 
Present work & 4.0\ \ \  & 3.6\ \ \  \\ 
\hline 
\cite{Tjon} & 4.09\ \  & 3.61\ \   \\
\hline 
\cite{levashov} & 4.23\ \  & 3.46\ \   \\
\hline 
\cite{fonseca}  & 3.905 & 3.597  \\
\hline 
\cite{belyaev} & 3.8\ \  \ & 4.9\ \  \  \\ 
\hline 
\cite{heiss} & 3.38\ \  & 3.25\ \  \\
\hline 
\cite{carbon} & 4.13\ \ & 3.73\ \  \\ 
\hline 
exp. \cite{exp} & 3.91 $\pm $ 0.12 & 3.6 $\pm $ 0.1 \\   
\hline 
\end{tabular}
\end{center}
\newpage 
\vskip 5cm 
\hspace*{50mm} Tab. 3\\
\begin{center}
\begin{tabular}{|c|c|c|c|c|c|}
\hline 
$S$    &   $A$     &$3+1$ phys.  &$3+1$ spur.&$2+2$ phys. & $2+2$ spur \\
\hline 
0      &   14.7    &  $+$       & $+$        &$+$          &$+$ \\
\hline 
0     &     15.4   &  $+$        & $-$        & $+$         &$-$ \\
\hline 
1     &     2.9    & $+$         & $+$        & $+$         & $+$ \\
\hline 
1    &      1.4    & $+$         & $+$        & $+$         & $-$ \\
\hline
1    &      1.2    & $+$         & $-$        & $+$         & $-$ \\ 
\hline 
\end{tabular}
\end{center}
\end{document}